\documentclass{ws-ijmpd}
\usepackage{epstopdf}
\usepackage{amssymb}                    
\usepackage{amsmath}
\usepackage{latexsym}
\usepackage{graphicx}
\begin{document}

\markboth{Prasia P. and Kuriakose V. C.}{Data Analysis of Massive
Gravitational Waves from Gamma Ray Bursts}

%
\catchline{}{}{}{}{}
%

\title{Data Analysis of Massive Gravitational Waves from Gamma Ray Bursts}
\author{P. Prasia\footnote{prasiapankunni@cusat.ac.in}\,\,  and V. C. Kuriakose\footnote{vck@cusat.ac.in}}
%

%
%
\address{Department of Physics, Cochin University of Science and
Technology, Kochi-682022, Kerala, India. } \maketitle

\begin{history}
\received{Day Month Year}
\revised{Day Month Year}
\end{history}
\begin{abstract}
We  investigate  the  detectability of massive mode of polarization
of Gravitational Waves (GWs) in $f(R)$ theory of gravity associated
with Gamma Ray Bursts (GRBs) sources. We obtain the beam pattern
function of Laser Interferometric Gravitational wave Observatory
(LIGO) corresponding to the massive polarization of GWs and perform
Bayesian analysis to study this polarization. It is found that the
massive polarization component with a mass of $10^{-22} eV/c^2$ is
too  weak to be  detected at LIGO with its current configuration.
\end{abstract}

\keywords{Gravitational Wave Detection, $f(R)$ Theory of Gravity,
Bayesian Analysis}
\ccode{PACS numbers:4.80.Nn, 95.55.Ym \and 04.30-w}
%
\section{Introduction}                                                                                          
General Theory  of Relativity (GTR) put forward by Einstein helps us
to look at the  Universe as  a   dynamic  system  amenable to
mathematical formulations  leading to  have  a   standard
cosmological  model of the  Universe. But  there   are   problems
with  the  standard cosmology based  on GTR,  both  at  the
conceptual  and  at   the cosmological/astrophysical levels\cite{1}.
The  latest  one  being the  observation  that   the  Universe  is
now  in  an accelerating phase. During  all  these  years,  attempts
are   being made  to modify  or  extend Einstein's  theory  to  find
explanations for the drawbacks  of  GTR.  Most  of  the  recent
studies  are all cosmological,  associated  with  or  replacing  the
constructs like \lq inflation\rq\,, \lq dark matter\rq\,  and \lq
dark energy\rq. All these studies fall  under  the  terminology \lq
Extended Theories of Gravity (ETG)\rq\, or \lq Alternative  Theories
of Gravity\rq\,\cite{2}. As a road to achieving modifications of GTR
for getting a correct explanation for the current astronomical
observations, it is better to consider a toy model as a tool to
explain the limitations of GTR and to test whether the  resulting
ETG form the right path to modifying GTR. $f(R)$ theories of
gravity\cite{2} form the simplest class of extended or modified
theories of gravity. Recently, massive gravity\cite{3,4} is
receiving great attention, as this model can be used to explain dark
energy  problem associated with  the late accelerating expansion of
the Universe.\\ \\
$f(R)$ theory comes out as a straight forward generalization of
Einstein-Hilbert action  for gravity  and is given by,
\begin{equation}
S=\frac{1}{2\kappa}%
{\displaystyle\int}
d^{4}x\sqrt{-g}f(R),
\end{equation}
where $\kappa=8\pi G,$ $G$ is the gravitational constant ($c=1$),
$g$ is the determinant of the metric tensor, $R$ is the Ricci scalar
and $f(R)$ is the generalization of $R$. $f(R)$ theory of gravity
makes a good toy model  for two reasons\cite{2}: $a)$ they are
sufficiently general to encapsulate some of the basic
characteristics of theories of gravity involving higher-order
curvature invariants, but at the same time they are simple enough to
be easy to handle and $b)$ they are unique among theories of gravity
involving higher-order curvature invariants, in the sense that they
seem to be the only ones which can avoid the long known and fatal
Ostrogradski
instability\cite{5}.\\ \\
The existence of Gravitational Waves (GWs) is a natural outcome of
GTR\cite{6}. With the path breaking discovery of GWs, supposed to be
from binary black hole merger\cite{7}, Laser Interferometer
Gravitational Wave Observatory (LIGO) serves as the center of
attention for future research in Gravitational Wave astronomy. A
second detection of GWs from the coalescence of two-stellar mass
black holes is also reported\cite{14}. LIGO has got three
specialized Michelson interferometers located at two sites $a)$
Hanford, $4$km-long H1 and $2$km long H$2$ detector $b)$ at
Livingston, a $4$km long L$1$ detector\cite{8}. Of the different GW
sources present, one of the most important classes that still lacks
a complete explanation is the Gamma Ray Bursts (GRBs). Studies are
going on and new developments are being made in understanding this
phenomenon. GRBs are intense flashes of $\gamma$-rays which occur
approximately once per day and are isotropically distributed over
the sky\cite{9,10}. Currently favored models of GRB progenitors are
grouped into two broad classes by their characteristic duration and
spectral hardness: $a)$ short GRB, the progenitors of which are
thought to be mergers of neutron star binaries or neutron-star black
hole binaries\cite{9a,9b} and $b)$ long GRB which are associated
with core-collapse supernovae\cite{9c}. Both mergers and supernovae
scenarios result in the formation of stellar-mass black holes with
accretion disk and the emission of GWs are expected in this process.
For the reasons mentioned above, GRBs form good sources of
gravitational radiation\cite{10a,10b}.\\ \\
Strong-field regimes form the best testing ground for the ETG and
GRBs provide a good strong-field regime for understanding
alternative theories of gravity. In the recent studies on the
discovery of GWs\cite{15a}, it is to  be noted that no studies were
done aiming at constraining parameters corresponding to any of the
alternative theories of gravity due to lack of predictions for what
the inspiral-merger-ringdown GW signal would look like in those
cases and no investigations were done for measuring the
non-transverse\cite{15b} components of GWs. All the facts throw
motivation for the modelling and parameter estimation of a GW event
occurring from GRB that is described by ETG. Testing of GWs in
alternative theories of gravity are discussed in general in the
literature\cite{15b,15,15c,15d,15e}. However, it is to be noted that
the production and detection of GWs on  the  basis of ETG for GRB
sources are not explored much.\\ \\
The detection of GWs involves the statistical analysis of the
observed data. It should tell whether the data contain the signal or
not or whether the data supports a certain theoretical model or not
with reliability. Statistical analysis can follow one of the two
perspectives: 1) Frequentist/Classical analysis 2) Bayesian
analysis. In a frequentist analysis, the probabilities are viewed in
terms of the frequencies of random repeatable events whereas
probabilities in Bayesian analysis provide a quantification of
uncertainty\cite{11}. From a Bayesian perspective, we can use the
machinery of probability theory to describe the uncertainty in model
parameters or in the choice of the model itself\cite{11a}. Bayesian
analysis can be \emph{parametric} or \emph{non parametric}. Non
parametric models constitute an approach to model selection and
adaptation where sizes of models are allowed to grow with data size
whereas in parametric models, a fixed number of parameters are
used\cite{11b}. A Bayesian formulation of non parametric problem is
non trivial since a Bayesian model defines the prior and posterior
distribution on a single fixed parameter space, but the dimension of
this parameter space in a non parametric approach changes with the
sample size\cite{12}.\\ \\
%
%
So, inspired by the recent observation of GWs, in this paper we make
an  attempt to study massive GWs from $f(R)$ theory of gravity, a
class of ETG. Also we  study the possibilities of detecting massive
polarization component of GWs emanating from GRBs from such a theory
at LIGO. The paper is organized as follows: In Section $2$, the
antenna response functions for the $f(R)$ theory of gravity are
found out and the beam pattern is figured for seven random GRB
candidates.  A Bayesian non parametric approach towards the signal
detection of the massive polarization from GRB is studied in Section
$3$. In Section $4$ massive GW signal from simulated data is
analyzed with the help of Bayes factor and the probable Signal to
Noise Ratio is calculated. Section $5$ concludes the paper.
\section{Response function of LIGO detectors towards massive gravitational waves}                                                                                       
The vacuum field equation of metric $f(R)$ gravity from $(1)$ is
given by,
\begin{equation}
f^{\prime}(R)R_{\mu\nu}-\frac{1}{2}f(R)g_{\mu\nu}-(\nabla_{\mu}\nabla_{\nu}-g_{\mu\nu}\square)f^{\prime}(R)=
0.
\end{equation}
Taking $f(R)$ to be of the specific form, $f(R)=R+\lambda R^{2}$,
where $\lambda$ is  a constant, the general solutions of this
equation are,
\begin{equation}
\bar{h}_{\mu\nu}={\displaystyle\int}A(k)\exp(-i(k.r-\omega t))dk,
\end{equation}
and,
\begin{equation}
\bar{h}_{\mu\nu}={\displaystyle\int}A(k)\exp(-i(k.r-\omega_{m}t))dk,
\end{equation}
where $\omega\neq\omega_{m}$\cite{17,18,19}.\\ \\
In GTR there are only two polarizations for gravitational radiation,
$+$ and $\times$. A scalar component of gravitational radiation in
Brans-Dicke theory has been proposed and the detection of such a
component has also been discussed in Ref. $[30]$. Recently it is
shown in Ref. $[31]$ that semi-classical effective field theory also
admits massless scalar GW solution in addition to conventional
polarization modes of GTR. The paper also discusses astrophysical
sources of scalar GWs. The utilization of metric $f(R)$ gravity
results in additional polarization  sates compared to the usual
polarization states, $+$
and $\times$ in GTR.\\ \\
In $f(R)$ theory, GWs can have a massive scalar mode besides the
usual transverse-traceless modes in GTR. Six polarization modes are
possible in $f(R)$ theories\cite{20}. A recent study\cite{21} shows
that in metric $f(R)$ theory in addition to the $+$ and $\times$,  a
breathing mode which goes along  with the $+$ and $\times$  modes
and a longitudinal scalar mode  which moves propagating along  the
direction  of propagation of  the GWs with a velocity less than the
velocity of light exist. But in Palatini formalism $f(R)$ theories
possess only the usual transverse-traceless modes as in GTR. GWs in
most of the extended theories of  gravity possess  more than the two
usual polarization modes. The detection of  GWs is  particularly  a
challenging  issue and  it may be capable of distinguishing the
different modes and may help us  to find  the correct formulation of
gravity. In this paper, we consider, only the case of massive scalar
polarization
component\cite{22}. \\ \\
The effect of GWs is to produce a transverse shear strain and this
fact makes the Michelson interferometer an obvious candidate for a
detector. When GWs pass through the detector, then one arm of the
detector gets stretched in one direction whereas the other arm gets
compressed. The dimensionless detector response function $h$ of an
interferometric detector is defined as the difference between the
wave induced relative length change of the two interferometer arms
and is computed from the formula given as\cite{24},
\begin{equation}
    h(t)=\frac{1}{2}\textbf{n}_{1}.[\tilde{H}(t)\textbf{n}_{1}]-\frac{1}{2}\textbf{n}_{2}.[\tilde{H}(t)\textbf{n}_{2}],
\end{equation}
where $\textbf{n}_{1}$ and $\textbf{n}_{2}$ are unit vectors
parallel to the arms $1$ and $2$ respectively and $\tilde{H}$ is the
three-dimensional matrix of the spatial metric perturbation produced
by the wave in the proper reference frame of the detector.\\ \\
Once a detector is built, it will be difficult to move it or even to
change it's orientation and hence the location and orientation of
detector will decide   how  the  detector  is  sensitive to
gravitational wave sources and likelihood of its detection. Hence,
the matrix $\tilde{H}(t)$ can be written as\cite{25},
\begin{equation}
    \tilde{H}(t)=M(t)H(t)M^{T}(t),
\end{equation}
where $H(t)$ is the spatial metric perturbation given by\cite{19},
\begin{equation}
    \left(
  \begin{matrix}
    h_{+}              &  {h_{\times}}         & 0 \\
    {h_{\times}}       &   -h_{+}       & 0 \\
    0                  & 0              & m^{2} h_{s}\\
  \end{matrix}
\right).
\end{equation}
$M$ is the three dimensional orthogonal matrix of transformation
from the wave cartesian coordinates to the cartesian coordinates in
the proper reference frame of the detector. $m$ is the mass of the
additional scalar polarization component of GW. If we follow  Ref.
$[33]$, $h_{+}$ can be taken as $h_{+}$ + $h^{b}$, where $h^{b}$ is
the breathing polarization mode and $-h_{+}=h_{+}-h^{b}$.\\ \\
From $(5)$, $(6)$ and $(7)$, we can write the response function as,
\begin{eqnarray}
    h(t)=F_{+}(t)h_{+}(t)+F_{\times}(t)h_{\times}(t)+F_{s}(t)(m^{2}h_{s}(t))\\
        =F_{+}(t)h_{+}(t)+F_{\times}(t)h_{\times}(t)+F_{s}(t)h(t)'
\end{eqnarray}
where $h(t)'=m^{2}h_{s}$  and  we  have  ignored $h^{b}$;
$F_{+}(t)$, $F_{\times}(t)$ and $F_{s}(t)$ are called beam pattern
functions. The beam pattern function, also called as response
function, determines the sensitivity of the detector towards an
incoming GW from a source.\\ \\
In order to express the beam pattern function in terms of right
ascension $(\alpha)$ and declination $(\delta)$ of the GW source, we
follow Jaranowski et al.\cite{25}. Accordingly, the matrix $M$ can
be represented as,\\ \\
\begin{equation}
    M=M_{3}M_{2}M^{T}_{1},
\end{equation}
where $M_{1}$ is the matrix of transformation from wave to detector
frame coordinates, $M_{2}$ is the matrix of transformation from
celestial to cardinal coordinates and $M_{3}$ is the matrix of
transformation from cardinal to the detector proper reference frame
coordinates.\\ \\
\begin{equation}
M_{1}=\left(
    \begin{matrix}
    A     &B
            &C\\
    D    &E
            &F\\
     G      &H
            &I
    \end{matrix}
\right),
\end{equation}
where,
\begin{flalign*}
A=&\sin\alpha\cos\psi-\cos\alpha\sin\delta\sin\psi,\\
B=&-\cos\alpha\cos\psi-\sin\alpha\sin\delta\sin\psi,\\
C=&\cos\delta\sin\psi,\\
D=&-\sin\alpha\sin\psi-\cos\alpha\sin\delta\cos\psi,\\
E=&\cos\alpha\sin\psi-\sin\alpha\sin\delta\cos\psi,\\
F=&\cos\delta\cos\psi,\\
G=&-\cos\alpha\cos\delta,\\
H=&-\sin\alpha\cos\delta,\\
I=&-\sin\delta,
\end{flalign*}
\begin{equation}
M_{2}=
    \left(
    \begin{matrix}
    \sin\lambda\cos(\phi+\Omega t)   &\sin\lambda\sin(\phi+\Omega t)
    &-\cos\lambda\\
    -\sin(\phi+\Omega t)    &\cos(\phi+\Omega t)    &0\\
    \cos\lambda\cos(\phi+\Omega t)   &\cos\lambda\sin(\phi+\Omega t)
    &\sin\lambda
    \end{matrix}
    \right),
\end{equation}
and,
\begin{equation}
M_{3}=\left(
  \begin{matrix}
    -\sin(\gamma+\zeta/2)   & \cos(\gamma+\zeta/2)      & 0 \\
    -\cos(\gamma+\zeta/2)   & -\sin(\gamma+\zeta/2)     & 0 \\
    0                       & 0                         & 1 \\
  \end{matrix}
\right),
\end{equation}
where $\lambda$ is the latitude of the detector's site, $\Omega$ is
the rotational frequency of earth in the units $1/(sidereal\quad
hours)$ and $\phi$ is a deterministic phase which defines the
position of the Earth in its diurnal motion at $t=0$. $\gamma$
determines the orientation of the arms of the detector with respect
to local geographical directions, $\zeta$ is the angle between the
arms of the interferometer. $\textbf{n}_{1}$ and $\textbf{n}_{2}$
have the coordinates,
\begin{equation} \textbf{n}_{1}=(1,0,0),
\textbf{n}_{2}=(\cos\zeta,\sin\zeta,0).
\end{equation}
The beam pattern functions can be found from $(5)$-$(13)$ and are
given by,
\begin{equation}
F_{+}=\frac{1}{2}[(S^{2}-T^{2})-(S\cos\zeta+v\sin\zeta)^{2}-\\
(T\cos\zeta+w\sin\zeta)^{2}],\\
\end{equation}
\begin{equation}
F_{\times}=
\frac{1}{2}\left[2ST\sin^{2}\zeta-\sin 2\zeta(Sw+Tv)^{2}-2vw\sin^{2}\zeta\right],\\
\end{equation}
\begin{equation}
\begin{split}
F_{s} &= \frac{1}{4} \sin\zeta [2 \cos(2 (\gamma +\zeta ))
\sin\alpha [2 \cos\delta^2 \cos(2 (\phi +\Omega t))
\sin\alpha \sin\lambda+\\
&\cos\lambda \sin{2 \delta } [-\cos(\phi +\Omega t
)+\sin(\phi +\Omega t)]]+\\
&\sin(2 (\gamma +\zeta )) [-2 \cos\lambda^2 \sin\delta^2+\sin\alpha
\sin{2 \delta } \sin{2 \lambda } [\cos(\phi +\Omega t)+\\
&\sin(\phi +\Omega t)]+\cos{\delta }^2 \sin{\alpha }^2 [2
\cos(\lambda )^2+(-3+\cos(2 \lambda )) \sin(2 (\phi +\Omega t
))]]],\\
\end{split}
\end{equation}
where,
\begin{eqnarray*}
a   &=& \sin\alpha\cos\psi-\cos\alpha\sin\delta\sin\psi,\\
b   &=& -\sin\alpha\sin\psi-\cos\alpha\sin\delta\cos\psi,\\
c   &=& -\cos\alpha\cos\delta,\\
d   &=& -\cos\alpha\cos\psi-\sin\alpha\sin\delta\sin\psi,\\
e   &=&  \cos\alpha\sin\psi-\sin\alpha\sin\delta\cos\psi,\\
f   &=& -\sin\alpha\cos\delta,\\
g   &=& \cos\delta\sin\psi,\\
h   &=& \cos\delta\cos\psi,\\
i   &=& -\sin\delta,\\
j   &=& \left[a\sin\lambda\cos(\phi+\Omega t)+d\sin\lambda\sin(\phi+\Omega t)-g\cos\lambda\right],\\
k   &=& \left[b\sin\lambda\cos(\phi+\Omega t)+e\sin\lambda\sin(\phi+\Omega t)-h\cos\lambda\right],\\
l   &=& \left[c\sin\lambda\cos(\phi+\Omega t)+f\sin\lambda\sin(\phi+\Omega t)-i\cos\lambda\right],\\
m   &=& -a\sin(\phi+\Omega t)+d \cos(\phi+\Omega t),\\
n   &=& -b\sin(\phi+\Omega t)+e \cos(\phi+\Omega t),\\
o   &=& -c\sin(\phi+\Omega t)+f \cos(\phi+\Omega t),\\
S   &=& -j\sin(\gamma+\zeta/2)+m\cos(\gamma+\zeta/2),\\
T   &=& -k\sin(\gamma+\zeta/2)+n\cos(\gamma+\zeta/2),\\
v   &=& -j\cos(\gamma+\zeta/2)-m\sin(\gamma+\zeta/2),\\
w   &=& -k\cos(\gamma+\zeta/2)-n\sin(\gamma+\zeta/2),\\
\end{eqnarray*}
In this paper we are only concerned with the response function of
the massive scalar component of the polarizations. The behavior of
the response function of this massive component with respect to the
azimuth angle can be plotted using (17). As examples we have chosen
the GRB instances given in Table $1$.
\begin{table}
\centering \textbf{Table $1$ :} GRB instances chosen for the
analysis
\begin{tabular}{|l|r|r|r|}
  \hline\hline
  Sl.No. & GRB Name & RA & DEC \\
  \hline\hline
  1 & 100206A & $3^{h}8^{m}40^{s}$ & $13^{0}10'$ \\
  2 & 100213A & $23^{h}17^{m}30^{s}$ & $42^{0}22'$ \\
  3 & 100216A & $10^{h}17^{m}03^{s}$ & $35^{0}31'$ \\
  4 & 100225B & $23^{h}31^{m}24^{s}$ & $15^{0}02'$ \\
  5 & 091223B & $15^{h}25^{m}04^{s}$ & $54^{0}44'$ \\
  6 & 100410B & $21^{h}16^{m}59^{s}$ & $37^{0}26'$ \\
  7 & 070201 & $0^{h}44^{m}21^{s}$ & $42^{0}18'$ \\
  \hline
\end{tabular}
\end{table}
The sources given in the table corresponding to $Sl.No. 1-3$ are
short GRBs taken from Table I, $Sl. No.4-6$ are long GRB taken from
Table II of Abadie et al.\cite{26} and GRB 070201 is taken from
Table $1$ of Abbott et al.\cite{26a}
\begin{figure}[h]
\begin{center}
\includegraphics[width=7pc]{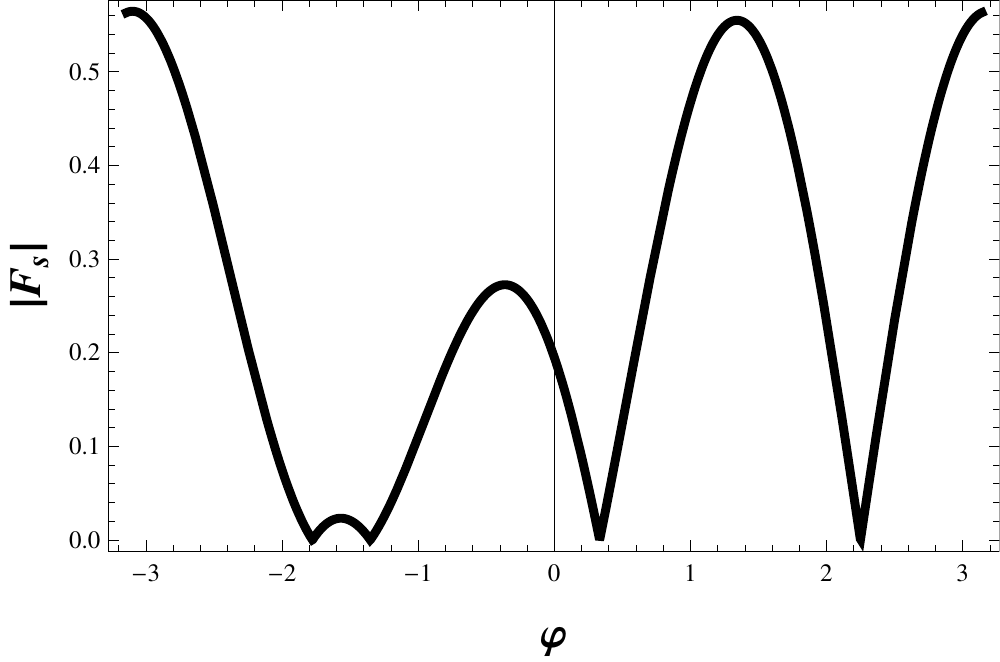}\includegraphics[width=7pc]{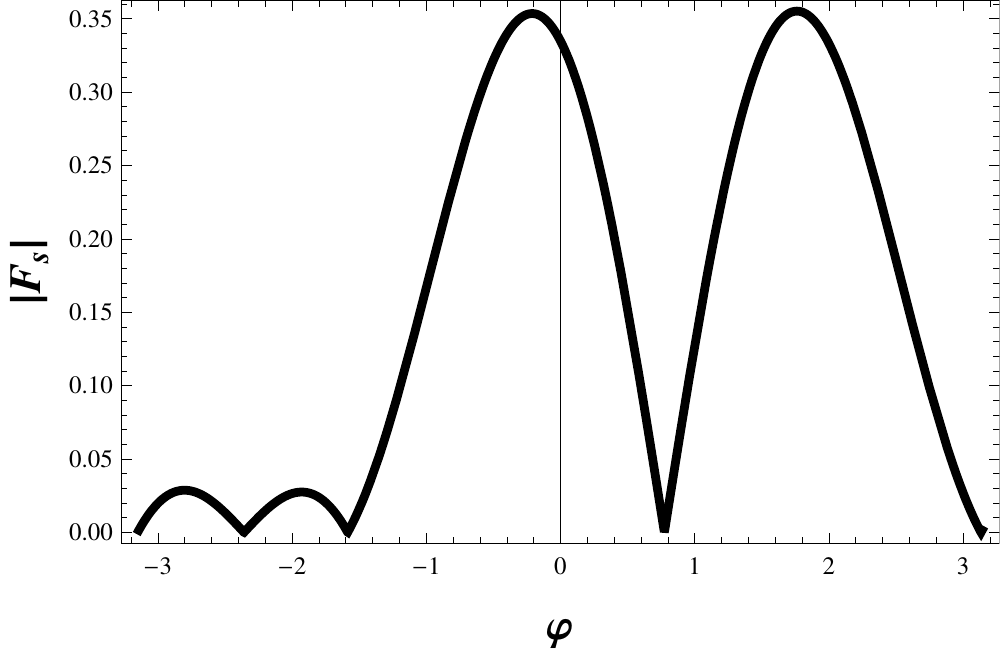}\hspace{0.7cm}
\includegraphics[width=7pc]{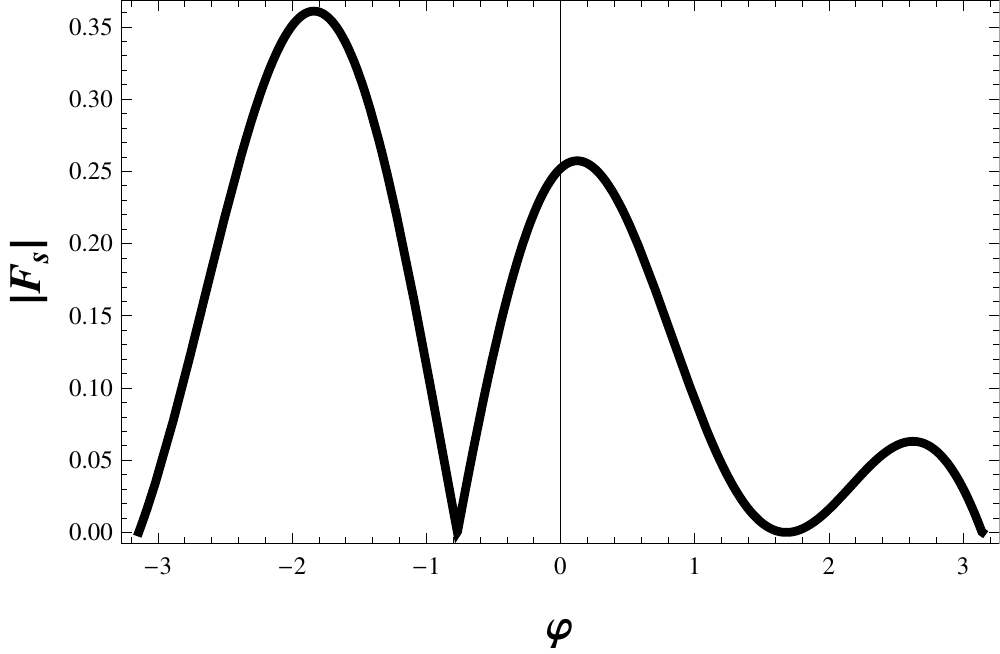}\includegraphics[width=7pc]{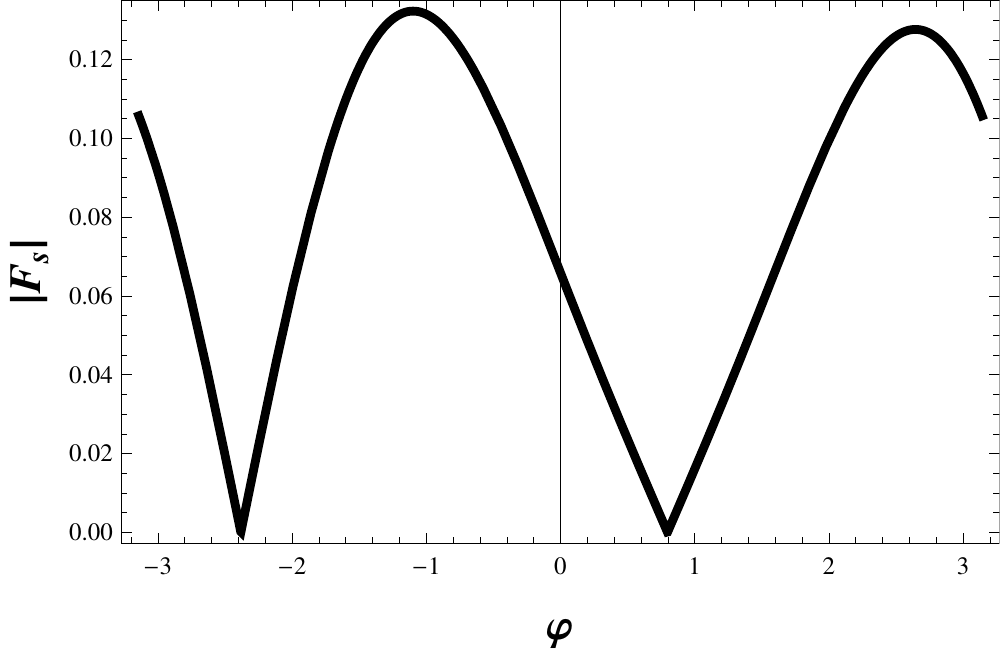}\\
\includegraphics[width=7pc]{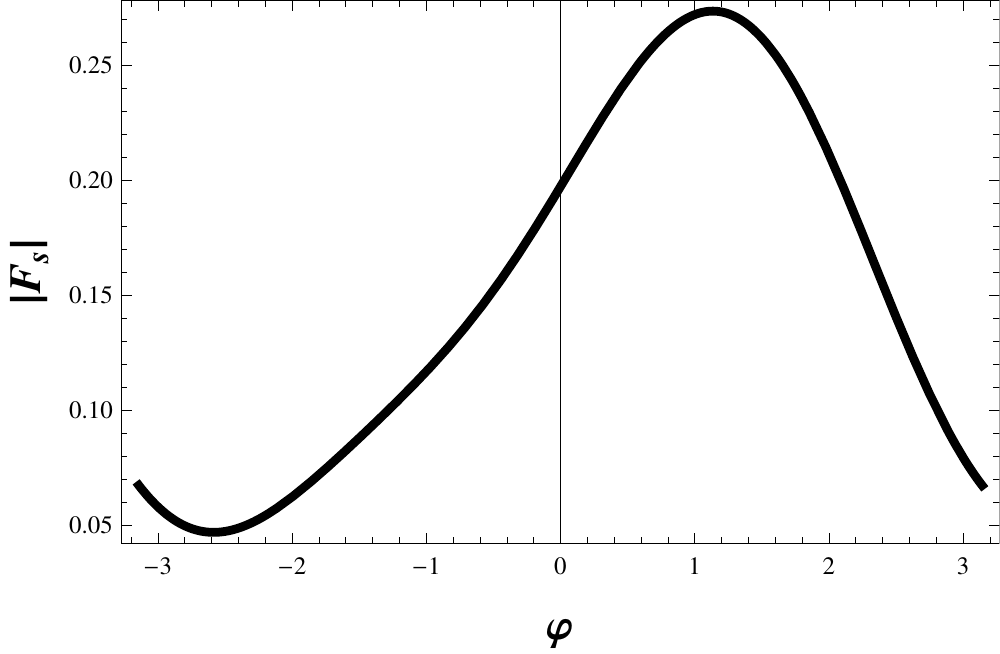}\includegraphics[width=7pc]{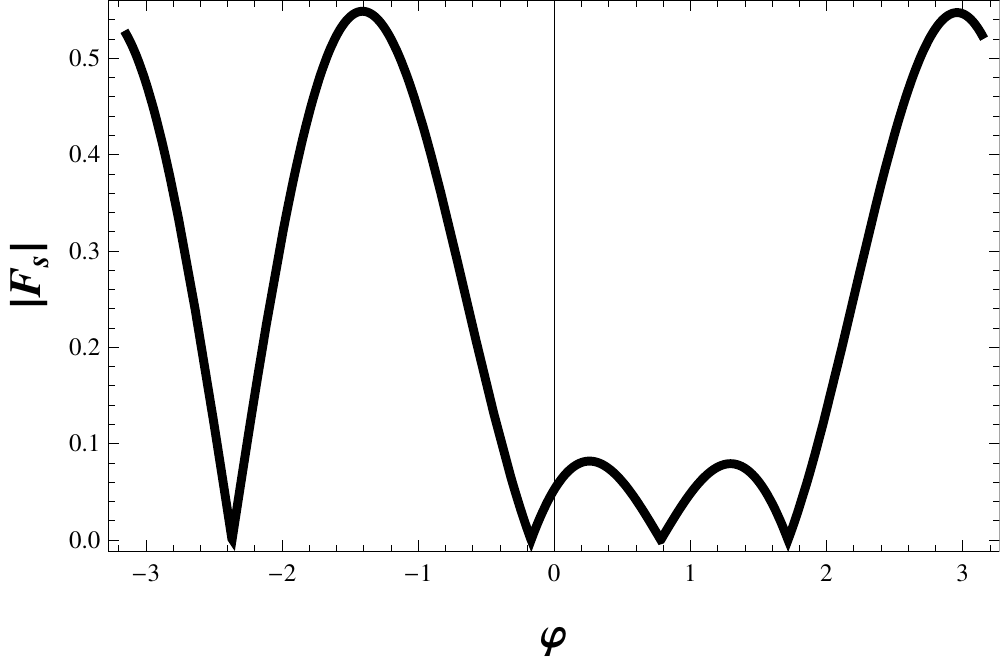}\hspace{0.7cm}
\includegraphics[width=7pc]{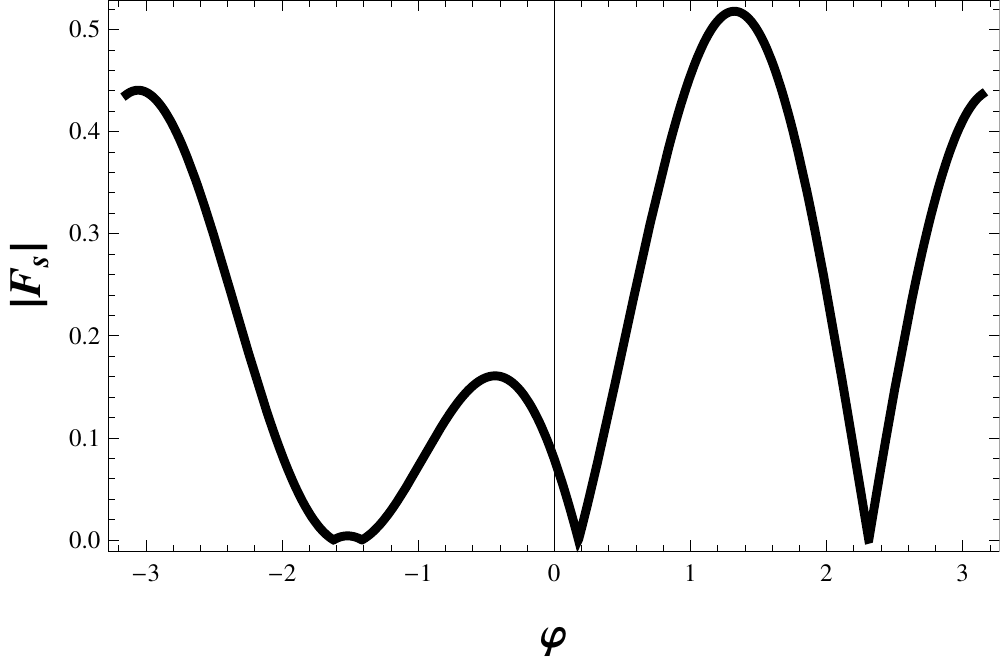}\includegraphics[width=7pc]{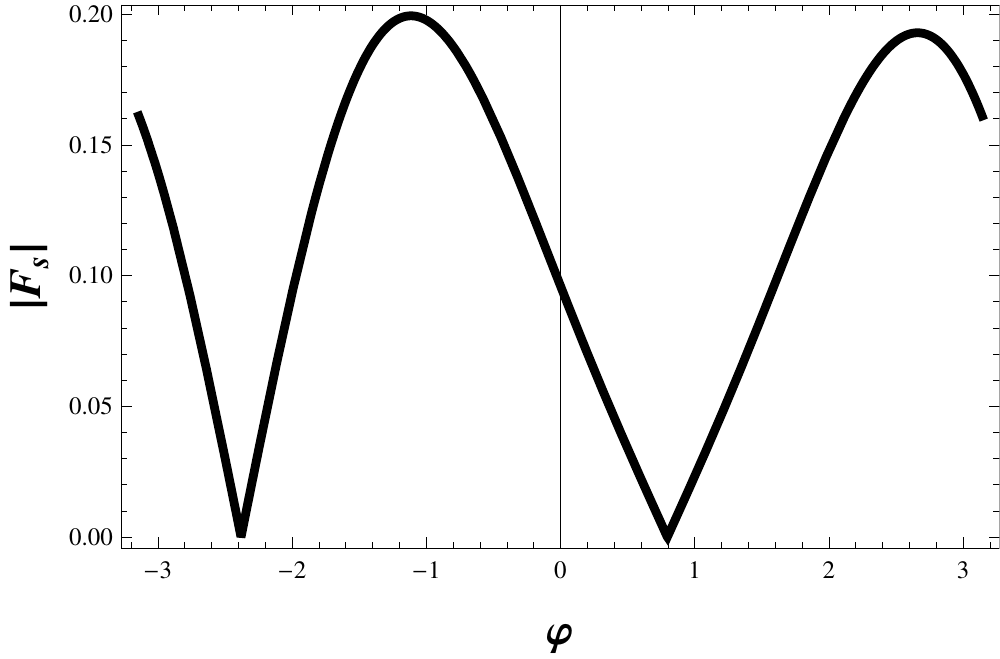}\\
\includegraphics[width=7pc]{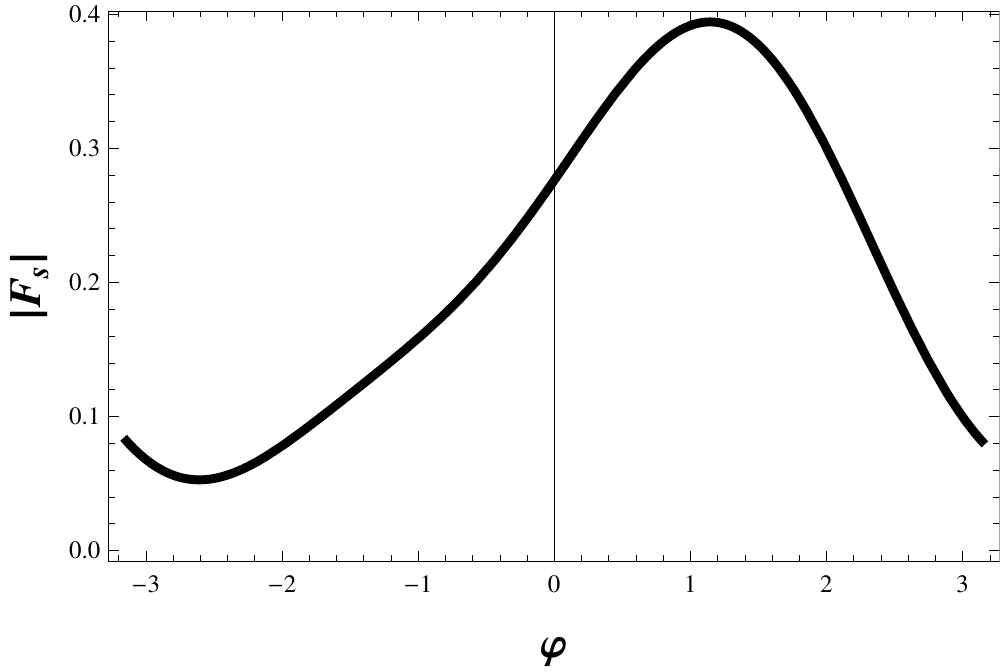}\includegraphics[width=7pc]{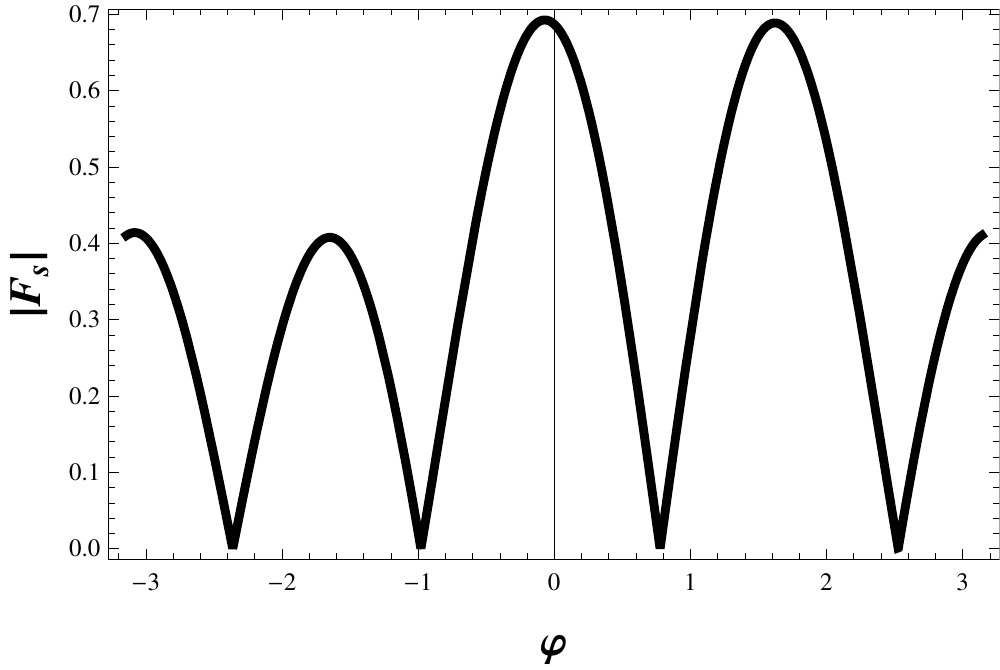}\hspace{0.7cm}
\includegraphics[width=7pc]{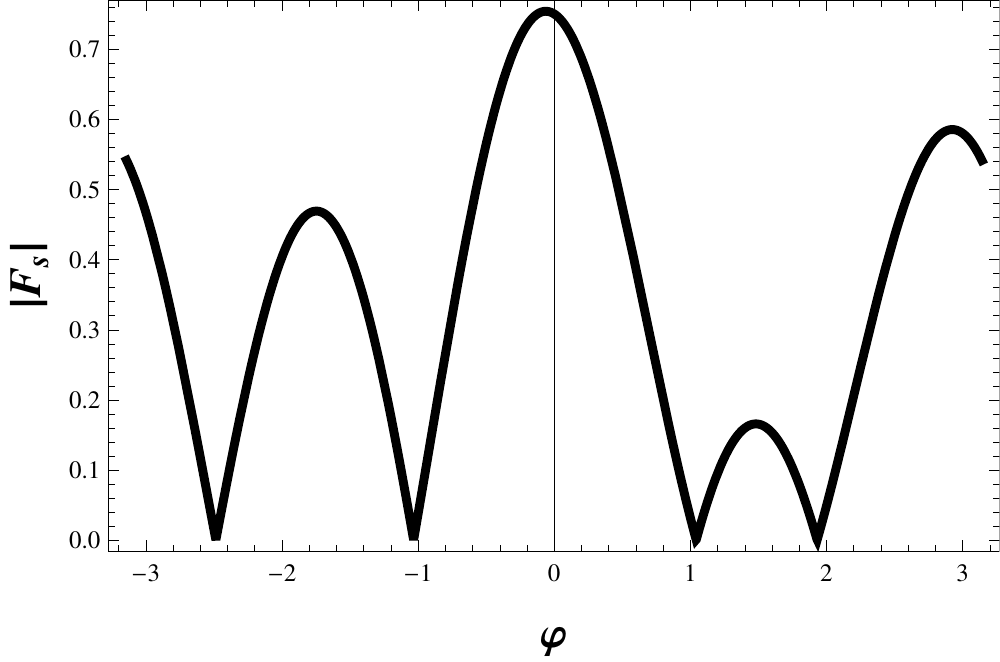}\includegraphics[width=7pc]{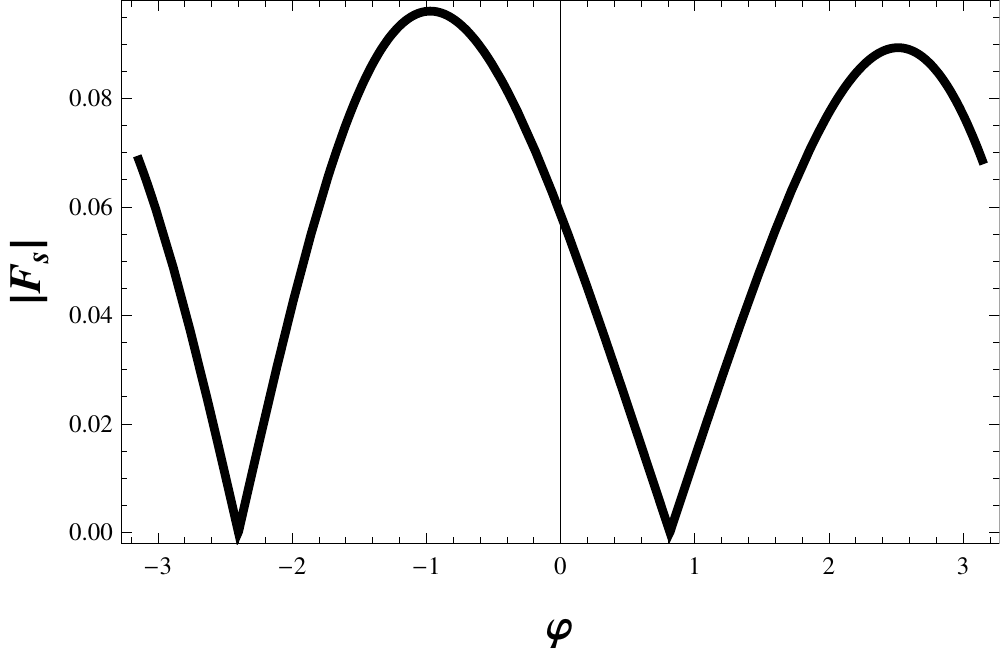}\\
\includegraphics[width=7pc]{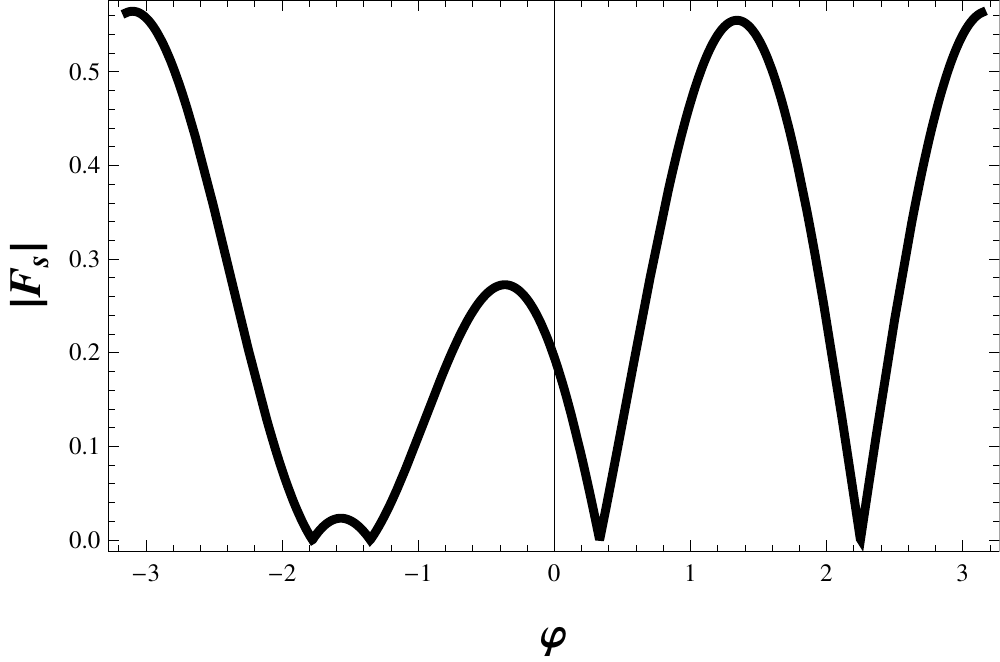}\includegraphics[width=7pc]{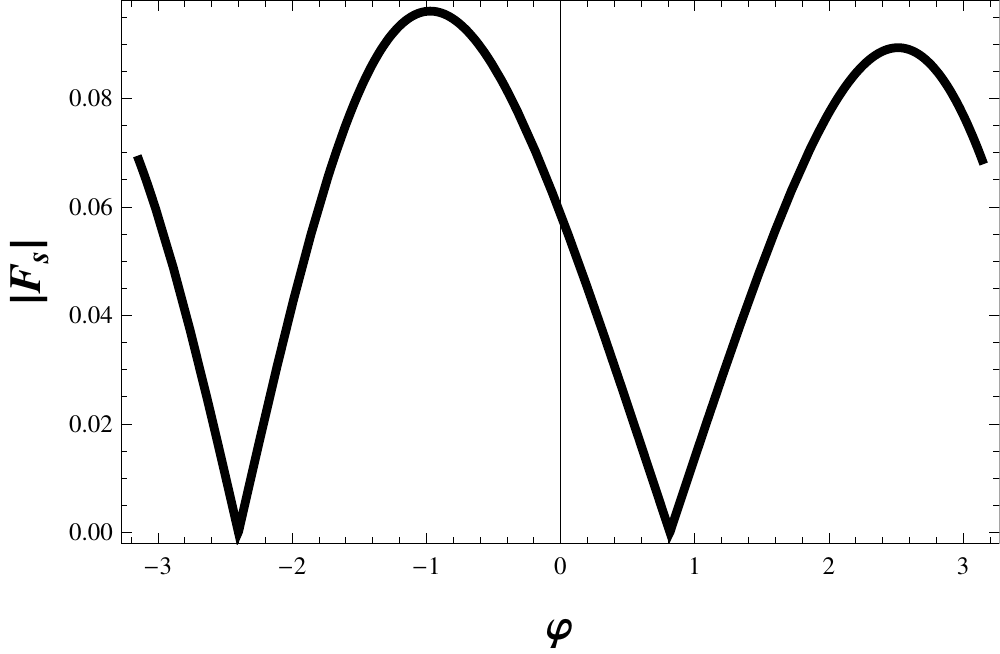}
\end{center}
\caption{Variation of $F_{s}$ with $\phi$ in the interval
$[-\pi,\pi]$ for the sources described in Table $1$ in their order,
for the LIGO Hanford (left) and LIGO Livingston (right)
respectively.}
\end{figure}
Fig. $1$ shows the variation of beam pattern function with $\phi$
for the above sources in the range $[-\pi,\pi]$ for the detectors
LIGO (Hanford) and LIGO (Livingston). From the figure, it can be
seen that for different sources the pattern function vary
differently, which means that depending on the location of the
detector, the response function changes. Also, the response
functions of the two LIGO detectors towards the massive component of
polarization of a GW for the same source, are found to be different.
\begin{figure}[h]
\begin{center}
\includegraphics[width=7pc]{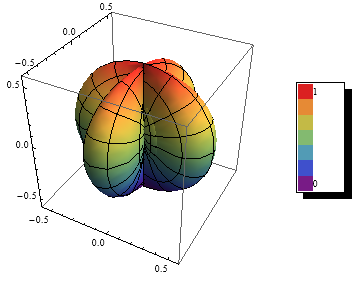}\includegraphics[width=7pc]{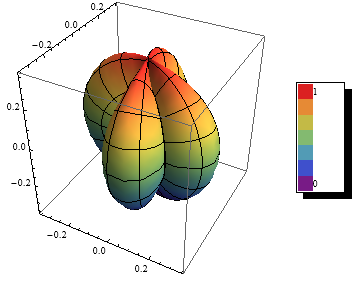}\hspace{0.7cm}
\includegraphics[width=7pc]{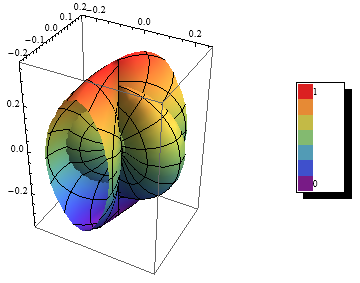}\includegraphics[width=7pc]{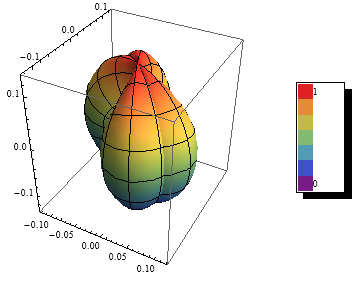}\\
\includegraphics[width=7pc]{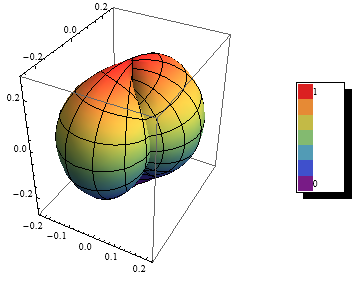}\includegraphics[width=7pc]{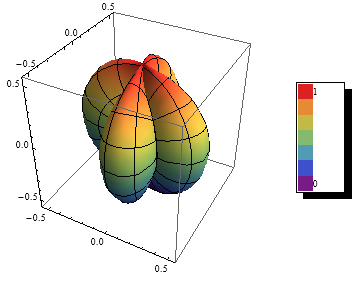}\hspace{0.7cm}
\includegraphics[width=7pc]{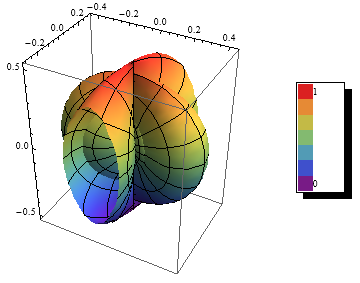}\includegraphics[width=7pc]{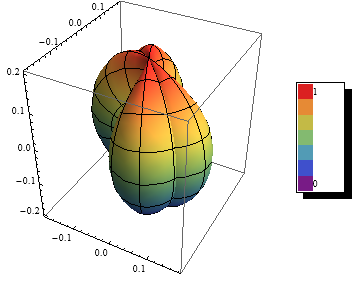}\\
\includegraphics[width=7pc]{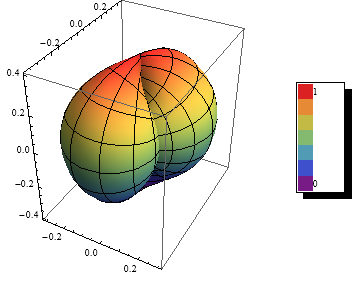}\includegraphics[width=7pc]{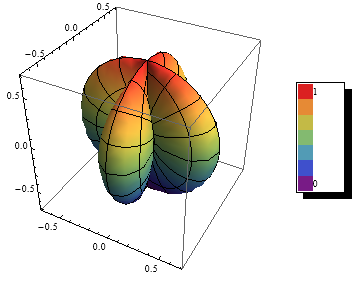}\hspace{0.7cm}
\includegraphics[width=7pc]{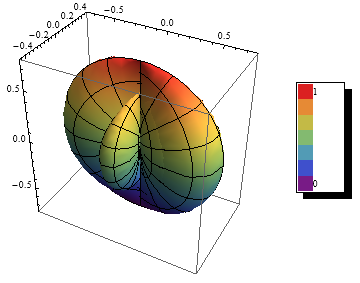}\includegraphics[width=7pc]{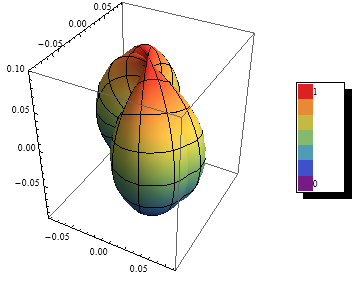}\\
\includegraphics[width=7pc]{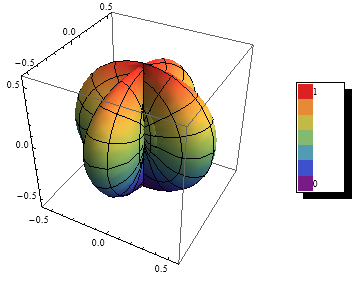}\includegraphics[width=7pc]{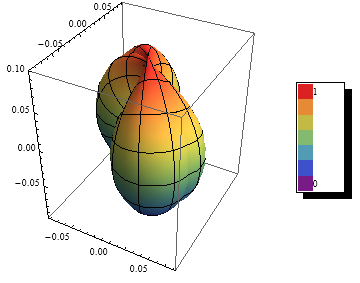}
\end{center}
\caption{Beam pattern function for the massive wave as a function of
$\phi$ and $\psi$ for the sources described in Table $1$ in their
order, for the LIGO Hanford (left) and LIGO Livingston (right)
respectively.}
\end{figure}
Fig. $(2)$ shows the beam pattern function behavior with the azimuth
angle $\phi$ and the polarization angle $\psi$. The antenna patterns
are in agreement with that proposed for massive scalar polarization
component\cite{23}. It can be easily inferred from the figure that
the beam pattern function behaves in a highly directional manner
towards an incoming wave of massive polarization which means that
the detector could detect a massive component of GW polarization
coming from a source only in a specific direction.
%
\section{A Bayesian approach to signal detection}                                                        
%
In this section, the Bayesian method is invoked to analyze the
massive scalar polarization of GW signal emanated from a GRB that
approaches the LIGO detector. Bayesian data analysis has already
been done for the case of Pulsar timing arrays for the $'+'$ and
$'\times'$ polarizations of GWs\cite{27}. The study of Bayes factor
as a norm for model selection, as to which model describes the data
best is also studied in this section.\\ \\
The Bayesian view is more general in which the probabilities provide
a quantification of uncertainty ie., the result of  Bayesian
analysis is a quantitative measure, stating how far the chosen
proposition is true. One advantage of the Bayesian viewpoint is that
the inclusion of prior knowledge arises naturally. Bayesian analysis
is completely controlled by the Bayesian law of conditional
probabilities that include the sum rule and the product rule. The
law is given by\cite{28},
\begin{equation}
    p(w|D)=\frac{p(D|w)p(w)}{p(D)},
\end{equation}
where $D$ is the observed data, $w$ is the parameter that defines
the proposition for $D$ and $p(w)$ is the prior probability. It is
the probability available before we observe the data. $p(w|D)$ is
called the posterior probability because it is the probability
obtained after we observe the data. $p(D|w)$ is the likelihood
function. It expresses how probable the observed data set is, for
different elements of the parameter vector \textbf{w}. $p(D)$ is the
normalization constant that makes the posterior distribution a valid
one and also ensure that it integrates to $1$. Then, $p(D)$ can be
written as,
\begin{equation}
        p(D)=\int{p(D|w)p(w)dw}.
\end{equation}\\
Applying the Bayesian approach to the GW signal analysis, we follow
Finn and Lommen\cite{27} to analyze the massive GWs from GRBs.
Suppose that the observed data is $D$ and let $h$ be the proposed
wave that describes the data $D$. The output data that we receive
from a detector will be a mixture of the original waveform $h$ and
the noise, $n$ of the detector, ie.,
\begin{equation}
  D = h(t)+n(t),
\end{equation}
where $h(t)$ is given by $(9)$. Here we deal only with the massive
scalar mode. Assuming that the wave exhibits only a single mode at a
time, the above equation can be written as,
\begin{equation}
    D=F_{s}(\theta, \phi)h_{s}+n(t).
\end{equation}
In this equation, we have taken $m=1$ for convenience. The noise is
assumed to be a zero mean additive Gaussian noise. Then, the
Bayesian law given by $(18)$ can be written in the form,
\begin{equation}
    p(h|D)=\frac{\Lambda(D|h)p(h)}{p(D)},
\end{equation}
where $p(h|D)$ is the posterior probability density, $\Lambda$ is
the likelihood function, $p(h)$ is the prior probability density and
$p(D)$, the normalization constant. The likelihood function
$\Lambda$ can be written as\cite{28}
\begin{equation}
    \Lambda(h|D)=N(\textbf{D}-\textbf{F}_{s}\textbf{h}_{s}|\textbf{C}),
\end{equation}
where $N$ denotes data drawn independently from a multivariate
Gaussian distribution. $\textbf{C}$ is the noise covariance and
$N(x|C)$, for a multivariate normal distribution with zero mean
random deviate $x$ given covariance C is given by,
\begin{equation}
    N(x|C)=\frac{exp{(-\frac{1}{2}x^{T}C^{-1}x)}}{\sqrt{(2\pi)^{\mathcal{N}}det||C||}}
\end{equation}
where $\mathcal{N}$ is the number of elements in vector $x$.
Assuming that the \emph{a priori} probability distribution is of
Gaussian form, we can write,
\begin{equation}
\begin{split}
    p(h_{s})&=N(h_{s}|\sigma_{s} I)\\
           &=[(2\pi \sigma_{s}^{2})]^{-1/2} exp{(-\frac{1}{2}\frac{h_{s}^{2}}{\sigma_{s}^{2}})},
\end{split}
\end{equation}
As already stated in Section $1$, the Bayesian non-parametric
formulation depends on the dimension of the parameter space.
Therefore, dimensionality should be included in the \emph{a priori}
distribution. The Gaussian distribution in higher dimensional space
containing many input variables is then given by\cite{27,29,30},
\begin{equation}
    p(h_{s})=[(2\pi \sigma_{s}^{2})^{\mathcal{N}}]^{-1/2}exp{(-\frac{1}{2}\sum_{k=1}^{\mathcal{N}}\frac{h_{k}^{2}}{\sigma_{s}^{2}})},
\end{equation}
where $\sigma_{s}$ is an undetermined constant, $\mathcal{N}$ can be
treated as the number of data taken and $I$ denotes an appropriately
dimensioned identity matrix. The normalization constant $p(D)$ is
the integral of the product of the likelihood function and the
\emph{a priori} probability density over all possible values of
$h_{s}$. Exploiting $(19)$, $(23)$, $(24)$ and $(25)$, we can write,
\begin{equation}
\begin{split}
    p(D)&=\int{\Lambda(h|D)p(h_{s})d^{\mathcal{N}}}h_{s}\\
    &=\frac{exp{(-\frac{1}{2}[\textbf{h}(t)^{T}\textbf{C}^{-1}\textbf{h}(t)])}}{\sqrt{(2\pi)^{\mathcal{N}}det||C||}}\\
    & \times\frac{exp{(\frac{1}{2}(F_{s}^{T}C^{-1}h(t))^{T}A^{-1}(F_{s}^{T}C^{-1}h(t)))}}{\sqrt{det||A||\sigma_{s}^{2 \mathcal{N}}}},
\end{split}
\end{equation}
where \textbf{A} can be expressed as,
\begin{equation}
    A=\sigma_{s}^{-2}I_{s}+F_{s}^{T}C^{-1}F_{s},
\end{equation}
and  $I_{s}$ is an appropriately dimensioned identity matrix.\\ \\
Finally, the posterior probability density $p(h|D)$ can be written
as,
\begin{equation}
p(h|D)=\sqrt{\frac{det||A||}{(2\pi)^{\mathcal{N}}}}exp{[\frac{1}{2}(h-h_{0})^{T}A(h-h_{0})]},
\end{equation}
where $h_{0}$ satisfies,
\begin{equation}
    A h_{0}= F_{s}^{T}C^{-1}h(t).
\end{equation}
It can be easily inferred from the above equation that $h_{0}$ is
the waveform that maximizes the probability density $p(h_{s}|h(t))$.
The amplitude Signal-to-Noise Ratio, $\rho$ associated with $h_{0}$
is given by
\begin{equation}
    \rho^{2}=(F_{s}h_{0})^{T}C^{-1}(F_{s}h_{0}).
\end{equation}\\
Finally, the quantity \emph{Bayes Factor} helps us to decide on
whether a signal is present or not. It chooses between different
models. For any observations D, the Bayes factor for $M_{1}$ against
$M_{0}$ is defined by\cite{31,32},
\begin{eqnarray}
  B_{10} &=& \frac{m_{1}(\theta)}{m_{0}(\theta)} \\
   &=& \frac{p(D|M_{1})}{p(D|M_{0})},
\end{eqnarray}
$\theta$ is some unknown parameter. The probability given in $(33)$
is nothing but the likelihood function. Therefore, employing the
form of $(23)$,
\begin{equation}
p(D|M_{1})=\Lambda(D|M_{1},h_{s},\sigma_{s}),
\end{equation}
gives the probability density of observations $D$ assuming the GW
signal described by parameter $\sigma_{s}$ is present,
\begin{equation}
    p(D|M_{0})=\Lambda(D|M_{0}),
\end{equation}
gives the probability density of $D$ assuming no signal is present.
The Bayes factor can then be written as\cite{27},
\begin{equation}
    B_{(D)}=\int{\frac{d^{2}\Omega_{k}}{4\pi}\frac{exp{(-\frac{1}{2}[\textbf{D}^{T}\textbf{C}^{-1}\textbf{D}])}}{\sqrt{det|A|\sigma_{s}^{2 dim
    h_{s}}}}}.
\end{equation}  \\
Now, from Bayes theorem, the posterior probability of model $M_{1}$
can be expressed through Bayes factor as\cite{31}
\begin{eqnarray}
    p(M_{1}|D)&=&
    \frac{p(M_{1})m_{1}(D)}{p(M_{1})m_{1}(D)+p(M_{0})m_{0}(D)}\\
    &=& \frac{p(M_{1})B_{10}}{p(M_{0})+p(M_{1})B_{10}},
\end{eqnarray}
where $p(M_{i})$ is the prior probability of model $M_{i}$ for
$i=0$. In the absence of any prior knowledge,
$p(M_{0})=p(M_{1})=1/2$. Therefore, the model $M_{1}$ is more likely
to be chosen if $p(M_{1}|D)>\frac{1}{2}$ or equivalently
$B_{10}>1$.\\ \\
Thus, Bayes factor is always positive. On the average, Bayes factor
will always favor the correct model. A Bayes factor large compared
to unity favor $M_{1}$ and a Bayes factor small compared to unity
favor the model $M_{0}$.
%
%
\subsection{Methodology}
Firstly, in order to check the  possibility of detecting massive GW
in the LIGO, the simulated data from $(21)$ is used. For that, a
simplest adhoc waveform given by a Gaussian distribution is used for
$h_{s}$, and can be written as in Abbott et al.\cite{33},
\begin{equation}
    h_{s}(t+t_{0})= h_{s,0}(\omega_{m})\cos{(2\pi f_{0}t)}\exp(-\frac{(2\pi
    f_{0}t)^{2}}{2Q^{2}}),
\end{equation}
where $t_{0}$ is the central time, $f_{0}$ is the central frequency,
which is taken in the range of $0$ to $200$Hz; $h_{s,0}$ is the
amplitude parameter that is characterized by $\omega_{m}$ of $(4)$
and is given as\cite{18,19},
\begin{equation}
    \omega_{m}=\frac{m}{\sqrt{1-v_g^{2}}},
\end{equation}
where $m$ is the mass corresponding to the additional scalar mode of
GW polarization, $v_{g}$ is the velocity of propagation of GW and
$Q$ is a dimensionless constant which represents roughly the number
of cycles with which the waveform oscillates more than half of the
peak amplitude. A standard choice in LIGO burst searches for $Q$ is
$8.9$. $t$ will be very short and is taken in the range $0$ to $1$s.
$h_{s,0}$ is given by\cite{10},
\begin{equation}
    h_{s,0}=\frac{1}{r}\sqrt{\frac{5GE_{GW}}{c^{3}Q f_{0} 4.
    \pi^{3/2}}}
\end{equation}
As an example to check whether massive scalar polarization resulting
from metric $f(R)$ gravity will be detected, we take the random
sample GRB$070201$. In order to simulate the detector output signal,
$(41)$ is substituted in $(39)$. This in turn is substituted in
$(21)$. For this candidate $E_{GW}=1.14\times 10^{-4}M_{\odot}c^{2}$
and $r=770$Kpc\cite{34}. Taking $F_{s}$ from $(17)$ and noise from
the LIGO Scientific Collaboration\cite{34a}, the simulated waveform
for different values of $m$ are shown in Fig. $4$.
\begin{figure}
  \includegraphics[width=14pc]{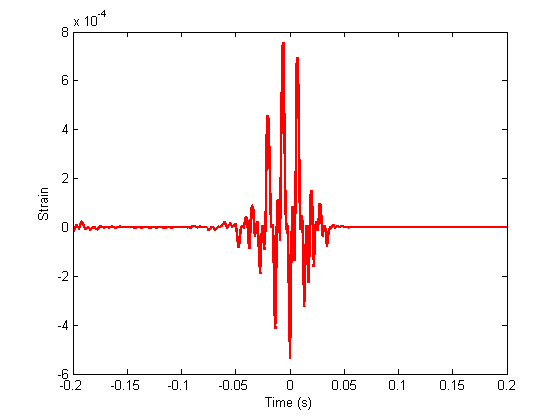}\hspace{0.5cm}\includegraphics[width=14pc]{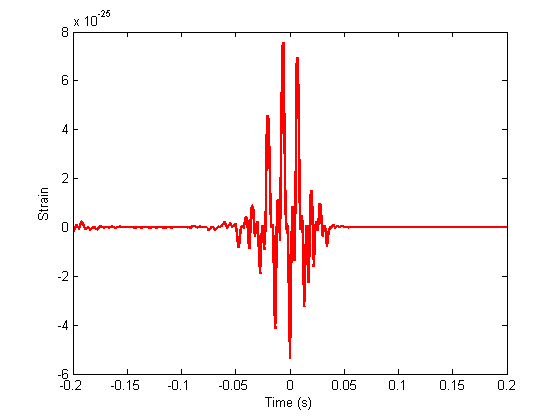}\\
  \includegraphics[width=14pc]{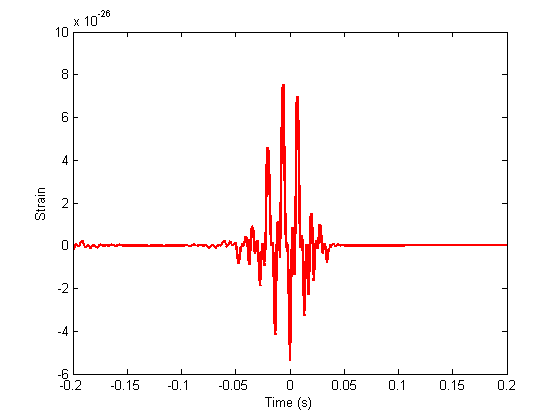}\hspace{0.5cm}\includegraphics[width=14pc]{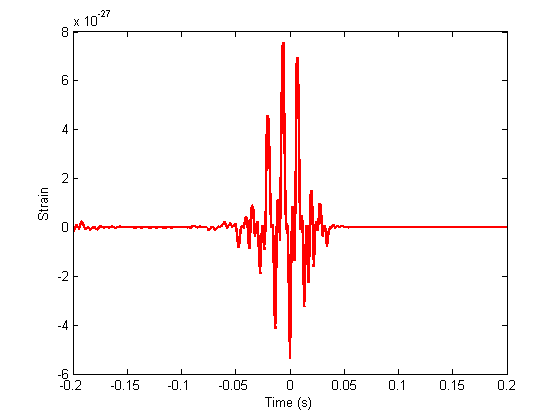}
  \caption{The simulated output signal for GRB070201 for $m=1$, $m=1\times 10^{-21}$,$m=1\times 10^{-22}$
  and $m=1\times 10^{-23}eV/c^2$ from top left}
\end{figure} \\
The data used in this section for Bayesian analysis is taken from
this simulated output waveform. Also, the predicted waveform is
taken from $(39)$. Even though it is not the wisest choice, it will
serve the purpose of the present work. After having the data $D$ and
the predicted waveform $h_{s}$, the parameter $\sigma_{s}$ is to be
estimated in order to get the most probable waveform and also the
Bayes factor. This is done by the optimization of $(27)$ with
respect to $\sigma_{s}$. While optimizing, maximization is used
since it is more favored\cite{35}. The equation as such is too
complicated to optimize and therefore it is simplified by taking the
logarithm of $p(D)$. This procedure is fully justified since
logarithm is a monotonically increasing function of its argument.
So, maximizing $\log{p(D)}$ with respect to $\sigma_{s}$ is
equivalent to maximizing $p(D)$. Thus $(27)$ becomes :
\begin{equation}
\begin{split}
    \log{p(D)}&=\log{[\frac{exp{-\frac{1}{2}[\textbf{h}(t)^{T}\textbf{C}^{-1}\textbf{h}(t)]}}{\sqrt{(2\pi)^{dim
    x}det||C||}}]}\\
    &+\log{[\frac{exp{\frac{1}{2}(F_{s}^{T}C^{-1}h(t))^{T}A^{-1}(F_{s}^{T}C^{-1}h(t))}}{\sqrt{det||A||\sigma_{s}^{2 dim
    h_{s}}}}]}.
\end{split}
\end{equation}
Since we are maximizing with respect to $\sigma_{s}$, the first term
on the right hand side can be suitably omitted as it is independent
of $\sigma_{s}$. The second term is optimized. Using the values of
matrix $\textbf{A}$ that is got by optimizing, $(30)$ is
simultaneously solved for $\textbf{h}_{0}$, the most probable
waveform (inferred waveform) for the given data. After getting
$\textbf{h}_{0}$, the signal to noise ratio $(\rho)$ can be
calculated using $(31)$. Then the Bayes factor can be evaluated
using $(36)$.
%
%
\subsection{Results}
The possible bounds on the mass of graviton corresponding to
different theories of gravity is discussed in the work of de Rham et
al.\cite{37a}. Bounds from the direct GW detection limit the
graviton mass as\cite{7} $m_g< 1.2\times 10^{-22}\,eV/c^2$. Invoking
these bounds, using the method discussed in the previous subsection,
the optimization is done by choosing the mass corresponding to the
additional scalar mode of GW polarizations as
$m=10^{-21},\,10^{-22}$ and $10^{-23}\,eV/c^2$. The results are
shown in Table $2$.
\begin{table}
\textbf{Table $2$ : }Results showing the calculation of log of Bayes
factor and SNR for different values of $m$.\\ \\
\begin{tabular}{|l|r|r|r|r|}
  \hline\hline
  $m (eV/c^2)$ & $\sigma_{s}$ & $\rho$ & $ln B_{D}$ & signal  \\
  \hline\hline
  $10^{-21}$ & $1\times 10^{-11}$ & $\sim0.1360$ & $\sim-1.7\times 10^{3}$ &
  weak/Absent\\
  \hline
  $10^{-22}$ & $1\times 10^{-11}$ & $\sim0.1360$ & $\sim-1.7\times 10^{3}$ &
  weak/Absent\\
  \hline
  $10^{-23}$ & $1\times 10^{-11}$ & $\sim0.1360$ & $\sim-1.7\times 10^{3}$ &
  weak/Absent\\
  \hline\hline
\end{tabular}
\end{table}
It can be seen that there is not much change in the values of Bayes
factor and SNR with the change in the mass of the scalar mode of
polarization in the range shown in the table. Comparing the values
of the Bayes factor and SNR given in Table $2$ with those in Table
$2$ of the Ref. $[39]$, it can be seen that the Bayes factor and SNR
values are very low indicating an absence of the signal. Thus it can
be concluded that with the given sensitivity and orientation of the
LIGO detector, a massive scalar polarization from $f(R)$ theory with
a value of mass in  the range $m=1\times 10^{-21}$ to $1\times
10^{-23}eV/c^2$  is unlikely to be detected. The comparison of most
probable waveform with the actual one is shown in Fig. $4$. This
null result can be compared with the results obtained in the works
of Aasi et al.\cite{36} and Xihao Deng\cite{37} where they predicted
null results for GRBs in the case of $+$ and $\times$ polarization
with the  existing observational set up.
\begin{figure}[h]
  \begin{center}
  \includegraphics[width=14pc]{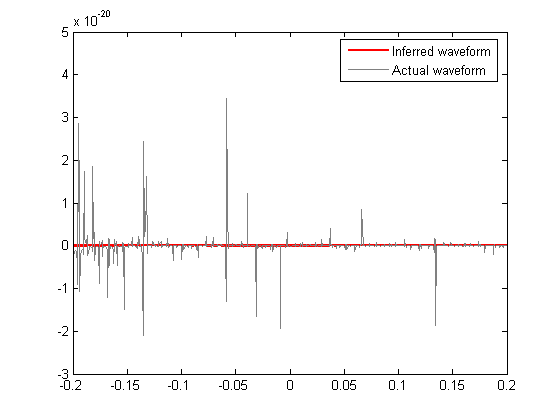}\\
  \end{center}
  \includegraphics[width=14pc]{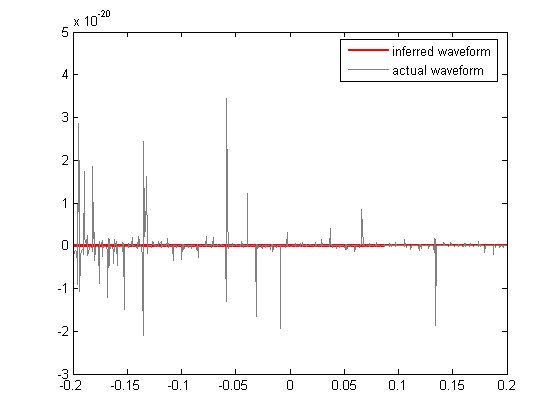}\hspace{.5cm}\includegraphics[width=14pc]{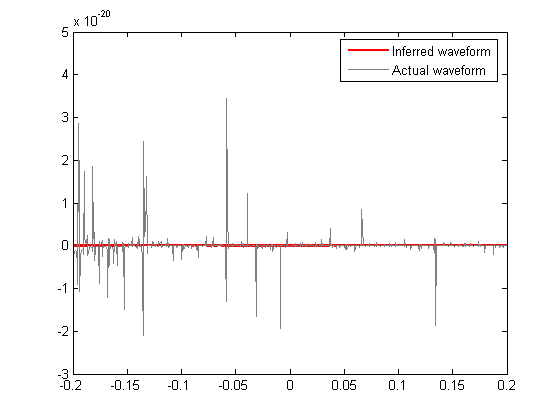}
  \caption{Comparison of the inferred and the actual waveforms for the LIGO detector for
  $m=1\times 10^{-21}$ on the top and for $m=1\times 10^{-22},\,1\times 10^{-23}$ respectively from left on the bottom row. }
\end{figure}
%
%
\section{Conclusion}                                                                                            
%
We have considered the production of massive GWs from a metric
$f(R)$ theory of gravity and the beam pattern it produces on an
interferometer detector. We have calculated the specific form for
the interferometer antenna response function in the detector
coordinates for the massive GWs. These are then considered for the
cases of LIGO Hanford and Livingston detectors for seven Gamma Ray
Burst (GRB) sources. These sources are selected at random. It is
found that the beam pattern functions are highly directional. They
are sensitive to the direction in which the wave comes. A Bayesian
analysis has been done to check the possibility of detecting  a
massive scalar component of GW polarizations from the source GRB
070201 using simulated data for LIGO, for the values of masses:
$m=10^{-21},10^{-22}$ and $10^{-23}eV/c^2$. The parameter of the
predicted waveform, which is nothing but the rms amplitude of the
wave, is determined by optimization method. The Bayes factor and the
SNR values are also determined. For all the cases the analysis gave
low values of SNR and Bayes factor. Thus with the model discussed in
this work for a GRB event and the beam pattern function, the massive
polarization is not likely to be detected. The results are prone to
change with a different \emph{a priori} waveform. Even though the
results presented in this paper are not conclusive enough, it gives
insight in to the study of GWs from alternative theories or extended
theories of gravity.\\
\section{Acknowledgements} The authors would like to thank the
reviewer for the comments for the improvement of the paper. One of
us (PP) would like to thank UGC, New Delhi for financial support
through the award of a Junior Research Fellowship (JRF) during the
period $2010$-$2013$. PP would also like to acknowledge Govt.
College, Chittur for allowing to pursue her research. VCK would like
to acknowledge Associateship of IUCAA, Pune.

\bibliographystyle{plainnat}

\bibliographystyle{spphys}
\end{document}